\documentstyle[sprocl]{article}

\begin{document}

\title{DEVIATIONS FROM LORENTZ INVARIANCE FOR ULTRAHIGH-ENERGY FERMIONS}
\author{R. E. ALLEN}

\address{Center for Theoretical Physics, Texas A\&M University\\
College Station, Texas 77843, USA\\ e-mail: allen@tamu.edu}

\maketitle\abstracts{In a new theory, local Lorentz invariance is a low-energy
symmetry which no longer holds when a fermion energy $E$ is well above 1
TeV. Here we find that the modified $E(p)$ relation is consistent with
observation, and is in fact nearly the same as in Einstein relativity. On
the other hand, there is a strong modification of the fermion equation of
motion and propagator at ultrahigh energy, which should lead to observable
effects.}

Recently a new fundamental theory was proposed.~\cite{a} One of its
predictions is a failure of Lorentz invariance at ultrahigh energies. Here
the implications of this feature will be considered in more detail.

\bigskip According to (4.4), (4.5), and (4.7) of Ref. 1, the Euclidean
Lagrangian density for a massless fermion has the form
\begin{equation}
\overline{{\cal {L}}}=\,\psi ^{\dagger }\left[ -\frac{1}{2M}h^{\mu \nu
}\partial _{\nu }\partial _{\mu }-i\left( \frac{1}{2}h^{\mu \nu }\partial
_{\nu }v_{\mu }+v_{\alpha }^{\mu }\sigma ^{\alpha }\partial _{\mu }\right)
\right] \psi
\end{equation}
where $M$ is a fundamental mass which is comparable to the Planck mass and
$h^{\mu \nu }=\delta ^{\mu \nu }$ is the metric in an initial (preferred)
coordinate system. The coupling to gauge fields and the extra fields of
(4.12) has been neglected. If the variation in $v_{\mu }$ is also neglected,
we obtain the Lorentzian Lagrangian density
\begin{equation}
{\cal {L}}=\,- \psi ^{\dagger }\left( -\frac{1}{2M}\eta ^{\mu \nu }\partial
_{\nu }\partial _{\mu } -iv_{\alpha }^{\mu }\sigma ^{\alpha }\partial _{\mu
} \right) \psi
\end{equation}
where $\eta^{\mu \nu }=diag(-1,1,1,1)$ is the Minkowski metric tensor.

\bigskip This is the generalization of (9.5) of Ref. 1, in which the first
term is neglected. In (4.15) of Ref. 1, $v_{\alpha }^{\mu }$ is interpreted
as the gravitational vierbein $e_{\alpha }^{\mu }$, which determines the
gravitational metric tensor $g_{\mu \nu }$ through the relations
\begin{equation}
g_{\mu \nu }=\eta_{\alpha \beta } e_{\mu }^{\alpha }e_{\nu }^{\beta }\quad
,\quad e_{\alpha
}^{\mu }e_{\nu }^{\alpha }=\delta _{\nu }^{\mu } \quad ,\quad e_{\alpha
}^{\mu }=v_{\alpha }^{\mu }.
\end{equation}
Then the expression for ${\cal {L}}$ yields the equation of motion
\begin{equation}
\left( \frac{1}{2M}\eta ^{\mu \nu }\partial _{\nu }\partial _{\mu
}+ie_{\alpha }^{\mu }\sigma ^{\alpha }\partial _{\mu }\right) \psi =0
\end{equation}
which can also be written
\begin{equation}
\left[ \frac{1}{2M}\left( -\,e_{\alpha }^{0}e_{0}^{\alpha }\partial
_{0}\partial _{0}+e_{\alpha }^{k}e_{l}^{\alpha }\partial _{l}\partial
_{k}\right) +ie_{\alpha }^{\mu }\sigma ^{\alpha }\partial _{\mu }\right]
\psi =0.
\end{equation}

\bigskip The cosmological model of Ref. 1 implies that
\begin{equation}
e_{\alpha }^{k}=\lambda \delta _{\alpha }^{k}\quad ,\quad e_{k}^{\alpha
}=\lambda ^{-1}\delta _{k}^{\alpha }=\lambda ^{-2}e_{\alpha }^{k}
\end{equation}
\begin{equation}
e_{\alpha }^{0}=\lambda _{0}\delta _{\alpha }^{0}\quad ,\quad e_{0}^{\alpha
}=\lambda _{0}^{-1}\delta _{0}^{\alpha }=\lambda _{0}^{-2}e_{\alpha }^{0}
\end{equation}
so (5) can be put in the form
\begin{equation}
\left[ \frac{1}{2M}\left( -\,\lambda _{0}^{-2}e_{\alpha }^{0}e_{\alpha
}^{0}\partial _{0}\partial _{0}+\lambda ^{-2}e_{\alpha }^{k}e_{\alpha
}^{l}\partial _{k}\partial _{l}\right) +ie_{\alpha }^{\mu }\sigma ^{\alpha
}\partial _{\mu }\right] \psi =0.
\end{equation}
After transforming to a locally inertial frame of reference, in which
$e_{\alpha }^{\mu }=\delta _{\alpha }^{\mu }$, we have
\begin{equation}
\left[ \frac{1}{2M}\left( -\,\lambda _{0}^{-2}\partial _{0}\partial
_{0}+\lambda ^{-2}\partial _{k}\partial _{k}\right) +i\left( \sigma
^{0}\partial _{0}+\sigma ^{k}\partial _{k}\right) \right] \psi =0.
\end{equation}
For a plane wave state $\psi \propto \exp \left( i\overrightarrow{p}\cdot
\overrightarrow{x}-iE\,t\right) ,$ this gives
\begin{equation}
\left[ \left( \,\beta E^{2}-\alpha p^{2}\right) +E-\overrightarrow{\sigma }
\cdot \overrightarrow{p}\right] \psi =0
\end{equation}
where $p^{2}=\overrightarrow{p}^{2}$ and
\begin{equation}
\alpha =\left( 2\lambda ^{2}M\right) ^{-1}\quad ,\quad \beta =\left(
2\lambda _{0}^{2}M\right) ^{-1}.
\end{equation}

\bigskip Now consider a particle with right- and left-handed fields coupled
by a mass $m$. The Lagrangian density is
\begin{eqnarray}
\cal{L} &=&\psi _{R}^{\dagger }\left( \frac{1}{2M}\eta ^{\mu \nu
}\partial _{\nu }\partial _{\mu }+ie_{\alpha }^{\mu }\sigma ^{\alpha
}\partial _{\mu }\right) \psi _{R} \\
&&+\psi _{L}^{\dagger }\left( \frac{1}{2M}\eta ^{\mu \nu }\partial _{\nu
}\partial _{\mu }+ie_{\alpha }^{\mu }\overline{\sigma }^{\alpha }\partial
_{\mu }\right) \psi _{L} \\
&&-m\psi _{R}^{\dagger }\psi _{L}-m\psi _{L}^{\dagger }\psi _{R}
\end{eqnarray}
where $\overline{\sigma }^{0}=\sigma ^{0}$ and $\overline{\sigma }
^{k}=-\sigma ^{k}$.~\cite{r} In a locally inertial frame, the resulting
equation of motion is
\begin{eqnarray}
\left[ \left(-\beta \partial _{0}\partial _{0}+\alpha \partial _{k}\partial
_{k}\right) +i\left( \sigma ^{0}\partial _{0}+\sigma ^{k}\partial
_{k}\right) \right] \psi _{R}-m\psi _{L} &=&0 \\
\left[ \left(-\beta \partial _{0}\partial _{0}+\alpha \partial _{k}\partial
_{k}\right) +i\left( \sigma ^{0}\partial _{0}-\sigma ^{k}\partial
_{k}\right) \right] \psi _{L}-m\psi _{R} &=&0
\end{eqnarray}
or, at fixed energy $E$ and momentum $\overrightarrow{p}$,
\begin{equation}
\left[ \left( \beta E^{2}-\alpha p^{2}\right) +E-\overrightarrow{\sigma }
\cdot \overrightarrow{p}\right] \psi _{R}-m\psi _{L}=0
\end{equation}
\begin{equation}
\left[ \left( \beta E^{2}-\alpha p^{2}\right) +E+\overrightarrow{\sigma }
\cdot \overrightarrow{p}\right] \psi _{L}-m\psi _{R}=0.
\end{equation}

\bigskip Solution of these coupled equations gives
\begin{equation}
\left( \beta E^{2}-\alpha p^{2}\right) +E=\pm \sqrt{p^{2}+m^{2}}
\end{equation}
or
\begin{equation}
2\beta E=-1\pm \left[ 1+4\beta \left( \alpha p^{2}\pm \sqrt{p^{2}+m^{2}}
\right) \right] ^{1/2}
\end{equation}
where the $\pm $ signs are independent. Then the lowest-order corrections to
Einstein relativity are given by
\begin{equation}
E\approx \sqrt{p^{2}+m^{2}}\left( 1-2\beta \gamma p^{2} + 2\beta^{2}m^{2}
\right) +\gamma p^{2}-\beta m^{2}\;\mbox{for}\;p\ll \alpha ^{-1},\beta ^{-1}
\end{equation}
where
\begin{equation}
\gamma \equiv \alpha -\beta.
\end{equation}
An argument given below implies that $\alpha =\beta$, or $\gamma =0$, but
let us first consider the behavior of ultrahigh-energy fermions in the more
general case.

\bigskip In particular, consider the GZK cutoff,~\cite{gzk,cg} which results
from collision of a charged fermion with a black-body photon. The incoming
photon has energy $\omega $ and momentum $(-\omega \cos \theta ,-\omega \sin
\theta ,0)$ in units with $\hbar =c=1$. The incoming fermion has mass $m_{a}$,
energy $E$, and momentum $(p,0,0)$. The outgoing fermion has mass $m_{b}$,
energy $E+\omega $, and momentum $(p-\omega \cos \theta ,-\omega \sin \theta
,0)$. If $\omega $ is small (as it is for a black-body photon), it is valid
to use
\begin{equation}
\Delta E=\frac{\partial E}{\partial p_{x}}\Delta p_{x}+\frac{\partial E}
{\partial p_{y}}\Delta p_{y}+\frac{\partial E}{\partial m^{2}}\Delta m^{2}
\end{equation}
with
\begin{equation}
\frac{\partial E}{\partial p_{k}}=v\frac{p_{k}}{p} \quad , \quad \frac
{\partial E}{\partial m^{2}}=\frac{1}{\left[ 1+4\beta p\left( 1+\alpha
p\right) \right] ^{1/2}}\frac{1}{2p}
\end{equation}
\begin{equation}
v=\frac{\partial E}{\partial p}=\frac{1+2\alpha p}{\left[ 1+4\beta p\left(
1+\alpha p\right) \right] ^{1/2}}
\end{equation}
for $m^{2}/p^{2}\ll 1$, so that
\begin{equation}
1+\frac{\partial E}{\partial p}\cos \theta =\frac{\partial E}{\partial m^{2}}
\frac{\Delta m^{2}}{\omega }
\end{equation}
and the threshold is for a head-on collision at the momentum $p$ satisfying
\begin{equation}
p\left( \left[ 1+4\beta p\left( 1+\alpha p\right) \right] ^{1/2}+\left(
1+2\alpha p\right) \right) =\frac{\Delta m^{2}}{2\omega }.
\end{equation}
Since the left-hand side increases with $p$, this condition will always be
satisfied, so there is still a GZK cutoff. This does not appear to be a
problem, however, since a recent analysis of the observations indicates that
the highest-energy protons are produced by neutral particles at much less
than cosmological distances.~\cite{fb}

\bigskip A violation of Lorentz invariance leads to the threat of
disagreement with precision experiments or high-energy observations,~\cite
{cg,ck} including prediction of new processes in the vacuum which are not
observed. An example is vacuum \v{C}erenkov radiation. Conservation of
energy and momentum implies that
\begin{equation}
-\omega =\Delta E=\frac{\partial E}{\partial p_{x}}\Delta p_{x}+\frac
{\partial E}{\partial p_{y}}\Delta p_{y}=\frac{\partial E}{\partial p}\left(
-\omega \cos \theta \right)
\end{equation}
so this process can occur if
\begin{equation}
v=1/\cos \theta \ge 1.
\end{equation}
For $m^{2}/p^{2}\ll 1$, the expression (25) implies that
\begin{equation}
v >1\;\mbox{for}\;\beta <\alpha \; \; ,\; \; v <1\;\mbox{for}\;\beta >
\alpha.
\end{equation}
If we were to have $\beta <\alpha$, the particle velocity at high momentum
would be greater than the velocity of light, and there would be a radiation
of photons in vacuum which is in conflict with observation.~\cite{cg}

\bigskip Next consider the process $photon \rightarrow e^{+}e^{-}$, which
will occur if
\begin{equation}
2E\left( p\right) =\omega =2p\cos \theta.
\end{equation}
It is reasonable to assume the threshold is below $\alpha ^{-1}$, so that
(21) is valid and the condition becomes
\begin{equation}
\sqrt{1+m^{2}/p^{2}}+\gamma p\approx 1+m^{2}/2p^{2}+\gamma p=\cos \theta .
\end{equation}
If we were to have $\beta >\alpha$, or $\gamma <0$, this condition would be
satisfied for $p^{3}>m^{2}|\gamma|^{-1}/2 $. Since observations indicate that
20 TeV photons do not decay in vacuum,~\cite{cg,cgw} $|\gamma|^{-1}$
would then have to lie above the Planck energy. This is inconsistent with
the theory of Ref. 1, where $\lambda $ and $\lambda _{0}$ are regarded as
small compared to unity.

\bigskip On the other hand, if $\beta =\alpha$, or $\gamma =0$, (21) reduces
to
\begin{equation}
E\approx \sqrt{p^{2}+m^{2}}\left( 1 + 2\alpha^{2}m^{2} \right)
-\alpha m^{2} \;\mbox{for} \;p\ll \alpha ^{-1}
\end{equation}
and the unphysical processes considered above do not occur. More
generally, (25) implies that the fermion velocity $v$ approaches the
speed of light $c=1$ as $m^{2}/p^{2}\rightarrow 0$, rather than
some different limiting velocity.

\bigskip The condition $\beta =\alpha $ can be understood if one makes the
following assumptions: (i) The vacuum energy density $\rho_{vac}$ is
not extremely large, in the sense that $|\rho_{vac}|\ll
m_{P}^{4}$, where $m_{P}$ is the Planck mass. (To be more specific,
this is the vacuum energy density which couples to gravity in the
present theory. It therefore omits the negative energy density 
from condensed Higgs fields.~\cite{a}) (ii) The dominant positive 
contribution to $\rho_{vac}$ comes from large-momentum positive-energy 
boson states, with velocity $c=1$. (iii) The dominant 
negative contribution to $\rho_{vac}$ comes from large-momentum
negative-energy fermion states, with velocity
$\overline{c}=\sqrt{\alpha /\beta }$.
(This is the velocity $v$ of (25) at large $p$.) (iv) At each large momentum
$p$, there are an equal number of these positive-energy and negative-energy
states, and they have the same momentum cutoff $p_{0}\sim m_{P}$.

\bigskip These general ideas may be illustrated with what is essentially a
toy model: The SO(10) gauge theory of Ref. 1 has 45 gauge bosons. If these
can be regarded as massless at the highest allowed momenta, each has 2
polarizations and an energy $\omega =p$. (For a cutoff frequency $\sim m_{P}$,
there is a ``Debye sphere'' with volume $\sim m_{P}^{3}$ and energy
density $\sim m_{P}^{4}$. If the
grand-unification scale is $\sim 10^{-3}\, m_{P}$, the
smaller sphere with this radius contributes only about $10^{-12}$
of the full energy, so even X-bosons can be
regarded as massless.) The total zero-point energy per momentum $p$ is then
$45\times 2\times \left( \omega /2\right) =45p$. On the other hand, there are
also the filled negative-energy states of (20). Suppose that only the 45
fermion fields of the standard model are relevant in the present context,
and that each of these has a single negative-energy state at each large
momentum $p$, with velocity $\overline{c}$. Their contribution to the vacuum
energy is then $-45\,\overline{c}\,p$, and the total for both fermions and
gauge bosons is $45\left( 1-\overline{c}\right) p$.

\bigskip In the cosmological model of Ref. 1, the value of $\alpha $ 
in the late universe is 
fixed by topology, since it corresponds to SU(2) rotations 
of the order parameter $\Psi_{s} $ in 3-space 
that are due to a cosmological instanton. (Recall that $\Psi_{s} $ describes an
SU(2) $\times $U(1)$\times $SO(10) GUT Higgs field which condenses in the
very early universe.~\cite{a}) The value of $\beta $, however, is not fixed
by topology and is free to vary, since it is associated with U(1)
oscillations of the order parameter in time. (A more detailed
cosmological picture is given elsewhere.~\cite{a1})
If the assumptions listed above
are valid, then the frequency of oscillation $v_{0}^{0}$ must adjust itself
so as to achieve a cancellation of the largest negative and positive
contributions to the vacuum energy. This means that
$\overline{c}=1$, or $\beta =\alpha$ .

\bigskip According to (33), we then obtain the standard relation of Einstein
relativity
\begin{equation}
E=\sqrt{p^{2}+m^{2}}
\end{equation}
except for very small terms involving the mass, so the present theory is
consistent with experimental tests of this relation. At ultrahigh energies
$p\sim \alpha ^{-1}=2\lambda ^{2}M$, however, the fermion equation of motion
is changed, and the propagator becomes $\left( -\alpha p^{\mu }p_{\mu
}+\rlap{/}p-m+i\epsilon \right) ^{-1}$ with the present conventions. This
change will affect the running coupling constants and other properties as
one approaches the GUT scale.

\section*{Acknowledgement}

This work was supported by the Robert A. Welch Foundation.


\section*{References}

\begin{thebibliography}{9}
\bibitem{a}  R.E. Allen, Int. J. Mod. Phys. A 12, 2385 (1997);
hep-th/9612041.

\bibitem{r}  P. Ramond, {\it Field Theory: A Modern Primer} (Addison-Wesley,
Redwood City, California, 1990), pp.18-19.

\bibitem{gzk}  K. Greisen, Phys. Rev. Lett. 16, 748 (1966); G.T. Zatsepin
and V.A. Kuz'min, JETP Lett. 41, 78 (1966).

\bibitem{cg}  S. Coleman and S.L. Glashow, hep-ph/9812418 and references
therein; Phys. Lett. B 405, 249 (1997).

\bibitem{fb}  G.R. Farrar and P.L. Biermann, Phys. Rev. Lett. 81, 3579
(1998).

\bibitem{ck}  D. Colladay and V.A. Kosteleck\'{y}, Phys. Rev. D 58,
116002 (1998), and earlier work referenced therein; hep-ph/9809521.

\bibitem{cgw}  J.W. Cronin, K.G. Gibbs, and T.C. Weekes, Annu. Rev. Nucl.
Part. Sci. 43, 883 (1993).

\bibitem{a1}  R.E. Allen, to be published in the proceedings of the
Cosmo-98 International Workshop on Particle Physics and the Early
Universe, edited by D. Caldwell (American Institute of Physics, New
York, 1999);astro-ph/9902042.

\end{thebibliography}
\end{document}